\documentclass[11pt]{article}
\usepackage{amsfonts,latexsym}
\newcommand{\cleqn}{\setcounter{equation}{0}}
\newcommand{\clth}{\setcounter{theorem}{0}}
\newcommand {\sectionnew}[1]{\section{#1}\cleqn\clth}
\newcommand{\beq}{\begin{equation}}
\newcommand{\eeq}{\end{equation}}

\newcommand{\beqa}{\begin{eqnarray}}
\newcommand{\eeqa}{\end{eqnarray}}
\newcommand{\beaa}{\begin{eqnarray*}}
\newcommand{\ben}{\begin{eqnarray*}}
\newcommand{\eaa}{\end{eqnarray*}}
\newcommand{\een}{\end{eqnarray*}}

\newcommand{\Nset}{\hfill\nonumber}
\newcommand{\text}{\textrm}
\newcommand \nc {\newcommand}
\nc \proof {\noindent {\em{Proof.\/ }}}
\nc \qed {$\Box$\hfill}
\newtheorem{theorem}{Theorem}[section]
\newtheorem{lemma}[theorem]{Lemma}
\newtheorem{proposition}[theorem]{Proposition}
\newtheorem{corollary}[theorem]{Corollary}
\newtheorem{definition}[theorem]{Definition}
\newtheorem{example}[theorem]{Example}
\newtheorem{remark}[theorem]{Remark}
\newtheorem{conjecture}[theorem]{Conjecture}
\newtheorem{question}[theorem]{Question}
\nc \bth[1] { \begin{theorem}\label{t#1} }
\nc \ble[1] { \begin{lemma}\label{l#1} }
\nc \bpr[1] { \begin{proposition}\label{p#1} }
\nc \bco[1] { \begin{corollary}\label{c#1} }
\nc \bde[1] { \begin{definition}\label{d#1}\rm }
\nc \bex[1] { \begin{example}\label{e#1}\rm }
\nc \bre[1] { \begin{remark}\label{r#1}\rm }
\nc \bcon[1] { \begin{conjecture}\label{con#1}\rm }
\nc \bque[1] { \begin{question}\label{que#1}\rm }
\nc {\eth} { \end{theorem} }
\nc {\ele} { \end{lemma} }
\nc {\epr} { \end{proposition} }
\nc {\eco} { \end{corollary} }
\nc {\ede} { \end{definition} }
\nc {\eex} { \end{example} }
\nc {\ere} { \end{remark} }
\nc {\econ} { \end{conjecture} }
\nc {\eque} { \end{question} }
\nc \eqref[1] {{\rm{(\ref{#1})}}}
\nc \thref[1]{Theorem \ref{t#1}}
\nc \leref[1]{Lemma \ref{l#1}}
\nc \prref[1]{Proposition \ref{p#1}}
\nc \coref[1]{Corollary \ref{c#1}}
\nc \deref[1]{Definition \ref{d#1}}
\nc \exref[1]{Example \ref{e#1}}
\nc \reref[1]{Remark \ref{r#1}}
\nc \conref[1]{Conjecture \ref{con#1}}

\def \a {\alpha}
\def \b {\beta}
\def \A {{\mathcal A}}

\def \O {{\mathcal O}}
\def \R {{\mathcal R}}

\def \Rset {{\mathbb R}}
\def \Cset {{\mathbb C}}
\def \Zset {{\mathbb Z}}
\def \Zset   {{\mathbb Z}}
\def \Nset {{\mathbb N}}

\def \ord { {\mathrm{ord}} }

\def \ad { {\mathrm{ad}} }
\def \Ad { {\mathrm{Ad}} }

\def \p { {\partial}}

\nc \Wr {Wr}
\nc \GRN { \Gr^{(N)} }
\nc \GRA[1] { \Gr_A^{(#1)} }   
\nc \GRAN { \GRA{N} }
\nc \GrA[1] { \Gr_A(#1) }\nc \GrAa { \GrA{\alpha} }
\nc \GRB[1] { \Gr_B^{(#1)} }   
\nc \GRBN { \GRB{N} }
\nc \GrB[1] { \Gr_B(#1) }
\nc \GrBb { \GrB{\beta} }
\nc \GRMB[1] { \Gr_{MB}^{(#1)} }   
\nc \GRMBN { \GRMB{N} }
\nc \GrMB[1] { \Gr_{MB}(#1) }
\nc \GrMBb { \GrMB{\beta} }

\begin{document}
\title{{\LARGE\bf{ Bispectral operators of prime order}}}
\author{
E. ~Horozov
\thanks{E-mail: horozov@fmi.uni-sofia.bg and @math.bas.bg}
\\ \hfill\\ \normalsize \textit{Institute of Mathematics and Informatics,}\\
\normalsize \textit{ Bulg. Acad. of Sci., Acad. G. Bonchev Str.,
Block 8, 1113 Sofia, Bulgaria }       }
\date{}
\maketitle
\begin{abstract}
The aim of this paper is to solve the bispectral problem for
bispectral operators whose order is a prime number. More precisely
we give a complete list of such bispectral operators. We use
systematically the operator approach and in particular - Dixmier
ideas on the first Weyl algebra. When the order is 2 the main
theorem is exactly the result of Duistermaat-Gr\"unbaum . On the
other hand our proofs seem to be simpler.

\end{abstract}
\setcounter{section}{-1}
\sectionnew{Introduction}

Bispectral operators have been introduced by F.A.Gr\"unbaum (cf.
\cite{G1,G2}) in his studies on applications of spectral analysis
to medical imaging.

 In the present paper we give complete classification of bispectral
operators of prime order. We start with some definitions and
 results that are needed to state our results, as well as to make
clear the connection  with other research.

An ordinary differential operator $L(x,\p_x)$ is called bispectral
if it has an eigenfunction $\psi(x,z)$, depending also on the
spectral parameter $z$, which is at the same time an eigenfunction
of another differential operator $\Lambda(z,\partial_z)$ now in
the spectral parameter $z$.  In other words we look for operators
$L$, $\Lambda$ and a function $\psi(x,z)$ satisfying equations of
the form:

    \beqa
    && L\psi=f(z)\psi, \label{1.1}\hfill\\
    && \Lambda\psi=\theta(x)\psi. \label{1.2} \hfill
      \eeqa

Although, as mentioned above, the study of bispectral operators
has been stimulated by certain problems of computer tomography,
later it turned out that they are connected to several actively
developing areas of mathematics and physics - the KP-hierarchy,
infinite-dimensional Lie algebras and their representations,
particle systems, automorphisms of algebras of differential
operators, non-commutative geometry, etc. (see e.g. \cite{BHY1,
BHY3, BHY4, BW, BW1, BW2, DG, K, W1, W2, MZ}, as well as the
papers in the proceedings volume of the conference in Montr\'eal
\cite{BP}).

In the fundamental paper \cite{DG} Duistermaat and Gr\"unbaum
raised the problem to find all bispectral operators and completely
solved it for operators $L$ of order two. The complete list is as
follows. If we present $L$ as a Schr\"odinger operator

$$L=(\frac{d}{dx})^2+u(x),$$ the potentials $u(x)$ of  bispectral
operators, apart from the obvious Airy ($u(x)=ax$) and Bessel
($u(x)=cx^{-2}$) ones, are organized into two families of
potentials   $u(x)$, which can be obtained by finitely many
"rational Darboux transformations"

(1) from $u(x)=0$,

(2) from $u(x)=-(\frac{1}{4})x^{-2}$.

Thus the classification scheme prompted by the paper \cite{DG} is
by the order of the operators. G. Wilson  \cite{W1} introduced
another classification scheme - by the rank of the bispectral
operator $L$ (see the next section for definitions).
 In the above cited paper \cite{W1} (see also \cite{W2}) Wilson
  gave a complete
description of all bispectral operators of rank 1 (and any order).
In the terminology of Darboux transformations (see \cite{BHY1})
all bispectral operators of rank 1 are those obtained by rational
Darboux transformations on the operators with constant
coefficients, i. e.  $L=p(\partial_x)\in \Cset$. In the above
mentioned papers \cite{DG, W1} the classification is split into
two, more or less independent parts. First, there is an explicit
construction of families of bispectral operators of a given class
(order 2 in \cite{DG}; rank 1 in \cite{W1}) The construction can
be given in terms of Darboux transformations of {\em "canonical"}
operators (a notion that needs clarification, see the last section
for some comments). A second part should be to give a proof that,
if an operator (in the corresponding class) is a bispectral one,
then it belongs to the constructed families.

In the last few years there has been increased activity
\cite{BHY1, BHY3, KRo, Z} in the the direction of constructing
classes of bispectral
 operators (\cite{BHY1, BHY3, KRo, Z}). For a survey on this
 subject, see  \cite{BP, H1} and the references therein.
 To the best of our  knowledge, all known up to
now families of bispectral operators can be constructed by
 the methods of \cite{BHY1,BHY3}. For a simplified exposition
 of these results, see the first part of \cite{H1}. A challenging problem is
 to prove that all the bispectral operators have already been found.
 A natural approach would be to divide the differential operators
 into suitable classes, e.g. - by
order as in \cite{DG} or by rank and to try to isolate the
bispectral ones amongst them. In \cite{HM} we have proposed
another classification scheme - that is to consider the
 operators with a fixed type of singularity at infinity.
The main result  of that paper is the classification of
 bispectral operators possessing
 the simplest type of singularity at infinity - the Fuchsian  one.
 My opinion is that all mentioned above classification schemes
may help each other as seen from the main results here.

In the present paper we return to the initial classification
scheme - that of \cite{DG}. We give a list of several families
that contains all bispectral operators whose order is a prime
number. Before stating the results we introduce some definitions
and notations which will be used also throughout the paper. We are
going to consider operators, normalized as follows:

\beq
L=\sum_{k=0}^{N} V_k(x)\p_x^k,\quad V_N=1,\quad V_{N-1}=0. \label{1.3}
\eeq
    It is well known that with the above normalization
 all the coefficients of $L$ are rational functions
 (see \cite{DG, W1} or the next section).

Now we can formulate the main result of the present paper.

\bth{1.1}
    An operator $L$, whose order is a prime number, is
bispectral if and only if it belongs to one of the following sets:

1. Generalized Airy operators:
 \beq A = \p^p + \sum_{j=1}^{p-2}a_j\p^j-x,\quad a_j\in\Cset;
\label{1.4} \eeq

2. Generalized Bessel operators:
 \beq
 B = x^{-p}(x\p-\b_1)\ldots(x\p-\b_p),\quad \b_j\in\Cset; \label{1.5}
\eeq

3. Operators with constant coefficients:

 \beq C = \p^p +  \sum_{j=1}^{p-2}a_j\p^j, \quad  a_j\in\Cset;
\label{1.6} \eeq

4. Operators, obtained by monomial Darboux transformations from
the Bessel operators having the property that at least one
difference $\b_i-\b_j \in p\Zset, i\neq j$;

5. Polynomial Darboux transformations from  operators with
constant coefficients.
     \eth

\bre{1.1}1. For $p=2$ this is just the content of the classical
result of Duistermaat-Gr\"unbaum \cite{DG}. Indeed, in that case
we have $\b_1-\b_2 \in 2\Zset$. In \cite{DG} this part of  the
theorem is formulated in a form close to this one. To obtain the
potential $u(x)=-(\frac{1}{4})x^{-2}$ mentioned above in 2) and
corresponding to $\b_1=\b_2= 1/2$ we have to perform monomial
Darboux transformation. Having in mind that a composition of
monomial Darboux transformations is again monomial darboux
transformation we get 2).  The potential $u(x)=0$ from 1)
corresponds to the only operator in 3., \thref{1.1} for $p=2$.

2. The different notions of Darboux transformations used here are
explained in the next section.

\ere

  \thref{1.1} is in fact a consequence of two
slightly more general results. Their importance lies in the fact
that they can be used for an induction process in further
classification (see \cite{H2}). In any case
 the  proofs do not simplify when restricted to  operators of prime order
  and seem to be more natural as performed here (see below).

At the end we briefly review the organization of the paper. In
Section 1. we recall some definitions and results, by now standard
for the problem, with the purpose to fix the notation and the
terminology. Section 2. treats the case of operators with bounded
(at infinity) coefficients $V_j$. We begin the section with an
auxiliary result. We show that from the normalized operator $L$ as
in \eqref{1.3} and satisfying only the so-called "$\ad$-condition"
with the polynomial $\theta$ , the function $f(z)$ can be chosen
naturally and then one can build the operator $\Lambda$  {\em
already normalized} only in terms of $L$ and $\theta$. This will
be needed after that to prove the following theorem for operators
with bounded coefficients:

 \bth{1.2} Let the rank of the bispectral operator L with bounded
 coefficients equals its order. Then it is a monomial Darboux transformation
  of a Bessel operator.
      \eth
    The essential part of the proof is to establish the vanishing of
    the coefficients $V_j$ at infinity. Then the result is contained
          in \cite{HM}.

In section 3. we consider operators with increasing coefficients.
We first  obtain a normal form for for their "leading terms"
(subsection 3.1).  Here we exploit once again (as in \cite{HM})
crucial ideas of Dixmier analysis of the first Weyl algebra
\cite{Dx}. When the order is a prime number we get the
non-vanishing (at infinity) part of $L$ to be (generalized) Airy
operator $A$. In the next subsection we develop some version of
wave (pseudo-differential) operators, expanded in negative powers
of Airy operators. This tool turns to be enough to prove (for
operators of any order, not only prime) in subsection 3.3 the
theorem:

\bth{1.3} Let the bispectral operator L = A +( vanishing at
infinity perturbation). Then the perturbation is zero.
           \eth
      We hope the careful reader has noticed some similarity with the
analysis in \cite{DG} for the latter class. We believe that our
proof is simpler and more transparent (and for this reason works
for higher order operators). One thing that we certainly benefited
from \cite{DG} is to realize that behind their calculations of the
normal form of second order operators there stands Dixmier
analysis on the first Weyl algebra $A_1$. On the other hand in
both steps we use different approaches.

The results of the present papers have been announced in the
second part of the survey paper \cite{H1}.

    {\bf Acknowledgements.} I am grateful to T.Milanov for useful
    conversations and for the fruitful collaboration in \cite{HM}.
    Important ideas in the present paper have their roots in
    \cite{HM}. The present version of the paper owes a lot to the
    referee. His constructive criticism has helped me to improve
 considerably the exposition - his suggestions made it clearer and
 corrected a number of errors. For all this I feel much obliged to
 him.     This paper has been partially supported by a grant
    No MM 1003-2000 of the National Fund "Scientific Researches"
    of the Ministry of Education of Bulgaria.

\sectionnew{Preliminaries}

In this section we have collected some terminology, notations and
results relevant for the study of bispectral operators. Our main
concern is to introduce unique notation which will be used
throughout the paper and to make the paper self contained. There
are also few results which cannot be found formally elsewhere, but
in fact are reformulations (in a form suitable for the present
paper) of statements from other sources.
\subsection{}
In this subsection we recall some definitions, facts and notation from Sato's
theory of KP-hierarchy \cite{S, DJKM, SW} needed in the paper.
For a complete
presentation of the theory we recommend also \cite{Di, vM}.
We start with the notion of {\it the wave operator} $K(x,\p_x)$.
This is a
pseudo-differential operator
\beq
K(x,\p_x)=1+\sum_{j=1}^{\infty}a_j(x)\p_x^{-j}, \label{2.1}
\eeq
with coefficients $a_j(x)$ which could be convergent or formal
power (Laurent) series.
In the present paper we will consider $a_j$ most often as formal
Laurent
series in $x^{-1}$. The wave operator defines the (stationary)
Baker-Akhiezer function $\psi(x,z)$:
\beq
\psi(x,z)=K(x,\p_x)e^{xz}. \label{2.2}
\eeq
From \eqref{2.1} and \eqref{2.2} it follows that $\psi$ has the following
asymptotic expansion:
\beq
\psi(x,z)= e^{xz}(1+\sum_1^{\infty}a_j(x)z^{-j}),\quad z\to {\infty}.
                                                                \label{2.3}
\eeq
Introduce also the pseudo-differential operator $P$:
\beq
P(x,\p_x)= K\p_xK^{-1}.                                         \label{2.4}
\eeq
The following spectral property of $P$, crucial in the theory
of KP-hierarchy,
is also very important for the bispectral problem:
\beq
P\psi(x,z)=z\psi(x,z).                                          \label{2.5}
\eeq
When it happens that some polynomial  of $P$, say $f(P)$, is a differential
operator,
we get that $\psi(x,z)$ is an eigenfunction of an ordinary
differential operator $L=f(P)$:
\beq
L\psi=f(z){\psi}. \label{2.6}
\eeq
It is possible to introduce the above objects in many different
ways, starting
with any of them (and with other, not introduced above). For us
it would be
important also to start with given {\it differential operator} $L$:
\beq
L(x,\p_x)= \p_x^N + V_{N-2}(x)\p^{N-2} + \ldots + V_0(x). \label{2.7}
\eeq
 One can define
the wave operator $K$ as:
\beq
LK=K f(\p). \label{2.9}
\eeq
An important notion, connected to an operator $L$ is the algebra
$\A_L$ of operators commuting with $L$ (see \cite{Kr, BC}). This algebra
is commutative one. The wave function $\psi(x,z)$ (defined in
\eqref{2.2}) is
a common wave function for all operators $M$ from $\A_L$:

\beq
M\psi(x,z)=g_M(z)\psi(x,z). \label{2.10}
\eeq
We define also the algebra $A_L$ of all functions $g_M(z)$ for which
\eqref{2.10} holds for some $M\in \A_L$. Obviously the algebras $A_L$ and
$\A_L$ are isomorphic.
Following \cite{Kr} we introduce {\it the rank of the algebra} $\A_L$ as
the greatest common divisor of the orders of the operators in $\A_L$.
\subsection{}
Here we shall briefly recall the definition of Bessel wave
function.
 Let $\beta \in
\Cset^N$ be such that

 \beq \sum_{i=1}^{N}\beta_i =
\frac{N(N-1)}{2}. \label{2.15} \eeq

 \bde{bess} \cite{F, Z, BHY1}
{\it {Bessel wave function}} is called the unique wave function
$\Psi_\beta(x,z)$ depending only on $xz$ and satisfying
 \beq
L_\beta (x, \partial_x) \Psi_\beta(x,z) =   z^N \Psi_\beta(x,z),
\label{2.17'2} \eeq
 where the Bessel operator $ L_{\b}(x,\p_x)$ is
given by \eqref{1.5}. \ede
   \noindent
      Because the Bessel wave
function depends only on $xz$, \eqref{2.17'2} implies

 \beqa &&D_x
\Psi_\beta(x,z)= D_z \Psi_\beta(x,z), \label{2.17'1}\\ &&L_\beta
(z, \partial_z) \Psi_\beta(x,z) =
   x^N \Psi_\beta(x,z).
\label{2.17'3}
\eeqa

    Next we define  {\em monomial} and {\em polynomial}
     Darboux transformations of Bessel operators. The definitions
     are  slight modification of the definitions
 given in \cite{BHY1}. Let $h(L_{\b})$ be a polynomial in a Bessel
operator.

 \bde{dt} We say that
the operator $\tilde{L}$ is {\em{polynomial\/}}
  {\em{Darboux transformation\/}} of $L_{\b}$
if there exist
 differential operators
$P(x,\partial_x)$, $Q(x,\partial_x)$ and a polynomial $h$ such
that

 \beqa
&&{h(L_{\b})}= QP \label{2.81} \\
 && \tilde{L}= PQ    \label{2.82}
  \eeqa
      and the operator $P(x, \partial_x)$  has the form

\beq P(x,\partial_x)=x^{-n}\sum_{k=0}^n p_k(x^N) D_x^k,
\label{2.84}
 \eeq
     where $p_k$ are rational functions, $p_n\equiv
1$. \ede

 \noindent
     We will use the following definition of
monomial Darboux transformations.

\bde{mon1} We say that the operator $ \tilde{L}$
 is a {\em monomial Darboux transformation\/} of the
Bessel operator $L_{\b}$ iff it is a polynomial Darboux
transformation  with $h(L_{\b})=L_{\b}^d$, $d\in\Nset$.
 \ede

   \bre{re}
  In \cite{DG} the authors work with {\em rational Darboux
   transformations}. It is easy to show that a composition of
   rational Darboux transformations is a monomial Darboux
   transformation.
   \ere

 We end this subsection by reformulating (in a weaker form)
the main results, which we need from \cite{BHY1, HM}.
\bth{2.11}
The polynomial Darboux transformations of the Bessel operators
are bispectral operators.
\eth
\bth{2.12}
If a bispectral operator L has vanishing coefficients at infinity,
it is a monomial Darboux transformation of a Bessel operator.
\eth
\subsection{}

Here we recall several simple properties of bispectral operators
following \cite{DG, W1}. As we have already mentioned in the
introduction we are going to study ordinary differential operators
$L$ of arbitrary order $N$ which are normalized as in \eqref{1.3},
i.e. with $V_N=1$ and $V_{N-1}=0$. Assuming that $L$ is bispectral
means that we have also another operator $\Lambda$, a wave
function $\psi(x,z)$ and two other functions $f(z)$ and
$\theta(x)$, such that the equations \eqref{1.1} and \eqref{1.2}
hold. The following lemma, due to \cite{DG}, has been fundamental
for all studies of bispectral operators.
    \ble{2.31} There exists a number $m \in \Nset$, such that

 \beq (\ad \, L)^{m+1}{\theta}=0.
          \label{2.31} \eeq
       \ele

     \noindent

         For its simple proof, see \cite{DG, W1}.
    To the best of my knowledge this is the only property of bispectral
    operators that is used in their studies.
    It is widely believed that the condition \eqref{2.31},
    called {\em the $\ad$-condition} is equivalent to bispectrality (provided
    \eqref{1.3} holds). In what follows we assume only that the
     $\ad$-conditon holds, i.e. we are not going to use the existence of $f(z)$,
$\Lambda$ and $\psi$. For us it would be important to construct
them only from $L$ and $\theta$ at least formally.
 This will be done in the next section.

 We will consider that $m$ is the minimal number
with this property. An important corollary of the above lemma is
the following result.

 \ble{2.32} Let the operator L be normalized
as in \eqref{1.3}. Then

(i)  The function $\theta(x)$ is a polynomial.

(ii) The coefficients $\a_j$ in the expansion \eqref{2.1} of the
wave operator $K$ are rational functions.
    \ele
     \proof We repeat
the simple proof following \cite{W1} as the proof introduces some
notions needed later. From the equation \eqref{2.31} it follows
that $$(\ad \, \p_x^N)^{m+1}(K^{-1}{\theta}K)=0.$$ On the other
hand the kernel of the operator $(\ad \, \p_x^N)^{m+1}$ consists
of all pseudo-differential operators whose coefficients are
polynomials in $x$ of degree at most $m$. This gives that
   \beq
  \theta(x) K=K\Theta, \label{2.32}
   \eeq
   with a pseudo-differential
operator $\Theta$ :
    \beq \Theta= \Theta_0
   +\sum_1^{\infty}\Theta_j\p_x^{-j} \label{2.33}
   \eeq
     whose
coefficients $\Theta_j$ are polynomials of degree at most $m$.
 We have $\theta(x)=\Theta_0(x)$.  This gives (i). Comparing the
coefficients at
$\partial_x^{-j}$ we find
that all the coefficients $\a_j(x)$ of $K$ are rational functions.
   \qed
   \bre{r1}
  We notice that at least one of the coefficients $\Theta_j$
 has degree {\it exactly m}, where $m$ from \leref{2.31} is minimal.
  This fact will be used later.
   \ere
    \noindent
The last lemma has as an obvious consequence one of the few general
results,
important in all studies of bispectral operators. Noticing
that the
coefficients of $L$ are polynomials in the derivatives of
$\a_j(x)$ we get
      \ble{2.34}
     The coefficients of $L$ are rational functions.
        \ele
    \bre{r2}
    Obviously the same results hold for the pair
     $\Lambda, f(z)$, when imposing
  the conditions \eqref{1.3} on $\Lambda$. But here we need to derive
  this statement from conditions {\em only on and $L$ and
  a suitable choice of $f(z)$} . This will be done in the next section.
         \ere

\sectionnew{Operators with bounded coefficients}
In this section we are going to prove our main result for operators with
 coefficients bounded near infinity.
As mentioned earlier we will consider slightly more general situation
 - operators for which the rank and the order coincide. The main part
of the proof is to show that in this case the coefficients are in
fact vanishing at infinity and hence the result follows from the
main theorem in \cite{HM}.

First I would like to fix the notation. Put

          \beq
    L= \p^p + \sum_{j=0}^{p-2}W_j(x)\p^j, \label{6.1b}
         \eeq
   where $W_j= c_j + V_j(x)$, $c_j$ are constants and
   $V_j(x)= \mathcal O(x^{-1})$.
       Next define the polynomial $f(z)$ to be :

      \beq
   f(z)= z^p + \sum_{j=0}^{p-2}c_jz^j. \label{6.2}
          \eeq
  In this way we can rewrite our operator in the form
    \beq
   L = f(\p) +  \sum_{j=0}^{p-2}V_j(x)\p^j, \label{6.1a}
     \eeq
      where $V_j(x)= \mathcal O(x^{-1})$.

Our first goal is to to show that starting with the normalized
bispectral operator $L$ the above choice of $f(z)$ leads to a
normalized operator $\Lambda$. The3 construction is formal, i.e.
the wave function is a formal series, but this is enough for what
follows. As all the auxiliary results (the three lemmas below) are
slight modification of corresponding results from \cite{HM} we
omit their proofs. It is well known that one can present the
operator $L$ in the form:

\beq
L = Kf(\p)K^{-1} \label{6.2a}
\eeq
  where the polynomial $f(z)$ is defined in \eqref{6.2}.
 In the next lemma, following \cite{DG} we find the simplest
restrictions on the coefficients of the wave operator $K$ and on $L$.

\ble{3.1}
(i) The coefficients $V_j(x),j=N-2,. . . 0$ of $L$ vanish at
$\infty$ at least as $x^{-2}$.

(ii) The coefficients $\a_j, \quad j=1, \dots$ of the wave operator $K$ vanish
at least as $x^{-1}$.
\ele

This lemma allows us to introduce following \cite{BHY4}  an
anti-isomorphism $b$ between   the algebra $\cal{B}$ of
pseudo-differential operators $P(x,\p_x)$ in the variable $x$ and
the same algebra $\cal{B}^{'}$ but in the variable $z$. More
precisely $\cal B$ consists of those pseudo-differential operators

    \ben
P=\sum_k^{\infty}p_j(x^{-1})\p_x^{-j},
    \een
  for which there is a number $n\in \Zset$ (depending on $P$) such
that all expressions $x^np_j(x^{-1}),\quad  j=k,k+1,\ldots$ are
formal power series in $x^{-1}$. The involution \ben b :{\cal B}
\longrightarrow {\cal B'} \een is defined by \beq
b(P)e^{xz}=Pe^{xz}=\sum_k^{\infty}
                   z^{-j}p_j(\p_z^{-1})e^{xz} ,
\quad
{\textrm{for }}
\quad P\in \cal B                                    \label{3.0}
\eeq
i. e. $b$ is just a continuation of the standard
anti-isomorphism between two copies of the Weyl algebra.
In what follows we will use also the anti-isomorphism
\beq
b_1:{\cal B} \longrightarrow {\cal B'},
\qquad    b_1(P)=b(\Ad _KP).                              \label{3.3}
\eeq
Obviously $b$ and $b_1$ can be considered as involutions of $\cal B$
and without any ambiguity we can denote the inverse isomorphisms
$b^{-1},b_1^{-1}:{\cal B'}\longrightarrow \cal B$
 by the same letters.

\noindent
    Since the operators $K$ and $\Theta= K^{-1}\theta K$ are from $\cal B$
we can {\em define} two operators $S$ and $\Lambda $ as follows:
     \beqa
      S(z,\p_z) & = & b(K(x,\p_x)),    \label{3.1}\\
\Lambda(z,\p_z) & = & b(\Theta).  \label{3.5}
     \eeqa
Explicitly one has
\beq
S=\sum_{j=0}^{\infty}z^{-j}\a_j(\p_z)
 =\sum_{j=0}^{\infty}a_j(z)\p_z^{-j},
  \qquad a_0=1                                      \label{3.2a}
\eeq
and also
\beq
\Lambda(z,\p_z)=\sum_{j=0}^{\infty}z^{-j}\Theta_j(\p_z)
               =\sum_{i=0}^{m}\Lambda_i(z)\p_z^i,
                                        \label{3.6}
\eeq where $\Lambda_m \neq 0$ (see \reref{r1}) and the
coefficients $\Lambda_i$ and $a_j$ should be viewed as {\em formal
power series}. As in \cite{HM} we can prove

     \ble{3.2} The coefficients $a_j$ of the operator
$S$ are rational functions.
        \ele

From the last lemma it follows that $\Lambda$ is normalized as
required  in \eqref{1.3}.
  Denote temporarily by $r$ the degree of the
polynomial $\theta$, i. e. if $\theta(x)=z^r+\ldots$.

\ble{3.3} With the choice of f(z) as in \eqref{6.2} the
coefficients $\Lambda_i$ of the operator $\Lambda$ are rational
functions and $\Lambda$ satisfies \eqref{1.2}.  The degree of
$\theta$ $r=m$ and \beq
  \Lambda_m=1,
\qquad \Lambda_{m-1}=0. \label{3.8} \eeq \ele

The point in the last lemma is that the normalization of $\Lambda$
is a consequence of the suitable choice of the polynomial $f(z)$.
Of course the wave function is only formal but this suffices for
the proof of the main theorem.

Now we are ready to give the classification of operators
 with bounded near infinity coefficients of operators with
 the same rank and order.

Let us fix the notation. Put

\beq
L= \p^N + \sum_{j=0}^{N-2}V_j(x)\p^j, \label{6.1}
\eeq
where $V_j= c_j + W_j(x)$, $c_j$ are constants and
$W_(x)= \mathcal O(x^{-1})$.
As before define the polynomial $f(z)$ to be :

\beq f(z)= z^N + \sum_{j=0}^{N-2}c_jz^j. \label{6.2b}
           \eeq
      With
this choice of $f(z)$ as we know (see ) the operator $\Lambda$ is
with rational coefficients. Now we are ready to prove the main
result of this section.

\bth{6.1}
   If the rank of the operator L with bounded at infinity
   coefficients equals its order $N$ then
all constants $c_j=0$, i.e. the coefficients $V_j =\O(x^{-1})$.
 \eth
              \proof We are going to use again the $\ad$-condition
\eqref{2.3}. Choose the number $m$ so that
$ad_{f(z)}^m(\Lambda)\neq0$ and
 $ad_{f(z)}^{m+1}(\Lambda)=0$. Simple computation shows that
\beq
ad_{f(z)}^m(\Lambda) =(-1)^m m!(f^{'}(z))^m
\label{6.3}
\eeq
On the other hand we have
      \beq
ad_L^m(\theta)= Q \neq 0 \label{6.4}
\eeq
       Obviously
\beq
[L,Q]=0. \label{6.5}
    \eeq
  This gives that $Q\in A_L$. From this and from the fact that the
rank of $L$ is equal to its order
$N$ we get that $Q$ is a polynomial in $L$:
\beq
Q= q_rL^r + q_{r-1}L^{r-1}+ \ldots,
\quad q_j\in\Cset, \label{6.6}
\eeq
where the coefficient $q_r\neq0$. Using the involution $b_1$ we
get

        \beq
 b_1(Q) = q_r f^r(z)+ \sum_{j=o}^{r-1}q_jf^j(z)
\label{6.7}
         \eeq
           Using \eqref{6.3} we have the following string of identities:
  \beq
b_1(Q)=b_1(ad_L^m(\theta))=(-1)^m(ad_{f(z)}^m(\Lambda))
 =  m!(f^{'}(z))^m.
\label{6.8}
   \eeq
      In this way \eqref{6.7} and \eqref{6.8} yield

\beq
 q_r f^r(z)+ \sum_{j=o}^{r-1}q_jf^j(z)=
  m!(f^{'}(z))^m.\label{6.9}
\eeq

Now we are going to compare both the degrees and the first two
coefficients of the two hand-sides of \eqref{6.9}. First notice
that comparing the degrees of the leading terms gives:

$$
(N-1)m=rN.
$$
This gives that
 $m$ is divisible by
$N$, i.e. $m=sN$ and $r=s(N-1)$. Next, comparing the coefficients
at the highest degree we get $q_r=m!N^m$. Suppose that some of the
coefficients $c_j$ of $f(z)$ are not zero . Denote the second
non-zero term after $z^r$ by $c_kz^k$. Computing the coefficient
at the second non-zero terms in both sides
 \eqref{6.9} we get $k=N-1$.
This is a contradiction to the normalizing condition $c_{N-1}=0$.
\qed

From the above theorem and from the main result in \cite{HM} (see
also Section 1.2, \thref{2.12}) we get the {\em proof} of
\thref{1.2}.

\sectionnew {Operators with increasing coefficients}

\subsection{Normal forms}
Let $L$ be a bispectral operator, normalized as in \eqref{1.3},
i.e.

     \beq
 L= \p_x^N + \sum_{j=0}^{N-2}V_j(x)\p_x^j.  \label{3.111}
        \eeq
We will consider that for some $j$ the corresponding
 coefficient $V_j$ is increasing at infinity, i.e.
 has Laurent expansion at infinity of the form:
 \beq
 V_j(x) =  \sum_{m=-\infty}^{r_j}a_{j,m}x^{m}, \label{3.2}
 \eeq
 where $r_j>0$ and $a_{j,r_j}\neq0$.
We call $r_j$ the order of $V_j$. In what follows we are going to
use several properties shared by the first Weyl algebra $A_1$ and
the larger algebra $\R[\p]$ of differential
 operators with rational coefficients.
Following \cite{Dx} we define filtration in $\R[\p]$. Let
 $\rho,\sigma \in \Rset$. We put $wt(x)=\rho$, $wt(\p)=\sigma$.
  Next the weight of $L$ is given by the following definition:

\bde{11.2} Assume that $L=V_n\p^n+V_{n-1}\p^{n-1}+\cdots +V_0$ is an
arbitrary element of $\R[\p]$. For each term $V(x)\p_x^i$ define its
weight \ben v_{\rho,\sigma}(V(x)\p_x^i)=\rho{(\ord V)}+\sigma{i} .
\een
Then the number

 \ben v_{\rho,\sigma}(L)  :=
                      \max_{0\leq i \leq n}
                         v_{\rho,\sigma}(V_i(x)\p^{i})
\een
will be called $(\rho,\sigma)$-order of $L$.
\ede
\noindent
The second definition associates to each differential
operator from ${\R}[\p]$ a $(\rho,\sigma)$-homogeneous polynomial.

\bde{11.3} Assume the notation of the previous definition and
denote by $I(L)$ the set $\lbrace i\in\lbrace
0,1,\cdots,n\rbrace\vert
v_{\rho,\sigma}(V_i\p^{i})=v_{\rho,\sigma}(L) \rbrace$. The
polynomial $f\in\Cset[x,x^{-1},y]$ defined as:

\beq f(x,y)=\sum_{i\in I}a_ix^{\ord V_i}y^i,
    \label{3.33} \eeq
where $a_i\in\Cset$ are uniquely determined from the expansion
\ben
  V_i=a_ix^{\ord V _i}+(lower\ order\ terms),
\een will be called polynomial associated with $L$. The operator
$L_0= :f(x,\p_x):$ will be called {\em the homogeneous part} of
$L$. Here, as usual the columns $::$ denote normal ordering, i.e.
the differentiation is pushed to the right.
 \ede
\noindent

 Consider again the bispectral operator \eqref{3.111}.
  In what follows we assume that $\rho$ and $\sigma$ are positive
integers. It is always possible to choose them in such a way that
the polynomial $f$ associated with $L$ has at least two terms of
the kind \beq
 f=y^N +\alpha x^k y^p + \ldots,\quad \alpha\neq0, \label{3.333}
 \eeq
where $k>0, \quad p\geq0$. For this purpose one can use the Newton
polygon. More precisely denote by $E(L)$ the set of points
$(m,j)$, such that $a_{m,j}\neq 0$, where $a_{m,j}$ is from
\eqref{3.2}. Consider the plane with the points of $E(L)$ and take
the convex closure of $E(L)$ (the Newton polygon). Then draw  a
line passing through the point $(0,N) \in E(L)$ and another point,
say $(k,j)\in E(L)$ with $k>0$ and such that the Newton polygon
remains below the line. This line is unique - all other points
$(k_1,j_1) \in E(L)$ with the same property lie on it. Then one
can find a non-zero solution in integers of the equation
$N\sigma=k\rho + j\sigma$. In our situation both $\rho$ and
$\sigma$ are positive as $j<N$ and at least one $V_j$ is
icreasing. Notice that with this choice the polynomial $f(x,y)\in
\Cset[x,y]$.

 Denote also by $g(x)=x^l$ the polynomial associated to
$\theta(x)$. Our goal is to find severe restrictions on the
polynomial $f$.
 We are going to use the following result which is
a slight modification of a particular case of the fundamental
Proposition 7.3 from \cite{Dx}.

          \ble{3.112} Suppose the element $L\in \R[\p]$ acts on a
          element $G$ nilpotently, i.e.

          $$ \quad  \ad_L^m(G)=0.  \quad m \geq1$$
   Let $f$ and $g$ be the polynomials associated to $L$ and $G$
   and $f$ contains at least two terms. If $\rho$ and $\sigma$ are
   positive integers and  $v_{\rho, \sigma}(L)>\rho + \sigma$, then
 one of the following cases holds:

(a)
 \beq
    f^s=g^r,  \label{3.3034}
  \eeq
  where $s$ and $r$ are the weights of $g$ and $f$,

     (b) $\sigma > \rho$, $\rho$ divides $\sigma$ and

     \beq
    f= X^n(X^m + \mu Y)^k,  \label{3.3044}
    \eeq

    (c) $\rho> \sigma$, $\sigma$ divides $\rho$ and
  \beq
   f=Y^n(Y^m + \mu X)^k  \label{3.3035}
 \eeq

 (d) $\rho=\sigma$ and

 \beq
   f=(Y+\lambda x)^n(Y + \mu X)^k  \label{3.3033}
 \eeq
            where $n\geq 0,\quad  k\geq 0, \quad m>0$
            and $\mu, \lambda\in\Cset$.

\ele

{\em Remarks on the  proof of the lemma}. The proof essentially
repeats that of lemma 7.3 from \cite{Dx}. We mention the minor
differences. While in \cite{Dx} all the polynomials
$(\rho,\sigma)$-associated with the elements of the Weyl algebra
belong to $\Cset[x,y]$ , here we work in $\Cset[x,x^{-1},y]$. The
fact that the polynomials belong to $\Cset[x,y]$ is needed in
\cite{Dx} mainly to speak  about their roots in some algebraic
closure of $\Cset(y)$ (respectively - $\Cset(x)$),when considered
as polynomials in $x$ (respectively in $y$). But the ring
considered here has the same property.

\ble{3.111}
 Suppose that $k>1$ ($k$ is from \eqref{3.333}). Then either
 $v_{\rho,\sigma}(L)>\rho + \sigma$ or $L$ cannot act nilpotently on $\theta$.
 \ele
 \proof
 Writing \eqref{3.333} in the form $f=y^p(y^{N-p}+\a x^k)+\ldots$ we
 notice that one can take $\rho=N-p$ and $\sigma=k\geq2$
 First suppose $N=k=2$ and $p=0$. Then $L$ cannot be nilpotent
 (see \cite{Dx}). We give a slightly different proof here.
 One can easily see  that if $ v_{\rho,\sigma}(\theta)=2l>0$, then
 $ v_{\rho,\sigma}ad_L(\theta)=2l$ unless
 $f^s=g^2$. But this is not possible as the polynomial
 associated with $\ad_L(\theta)$ contains a term with $y$ of degree
 one. Denote by $g_m$ the polynomial associated to $\ad_L^m(\theta)$.
  By induction on $m$ we see that the resulting
 applications of $ad_L$  always
contain terms where the power of $y$ is less than 2. This shows
that
 the equation  $f^s=g^2$ is impossible.

   Now we can suppose that either $p>0$ or $\max(N,k)\geq 3$ (we recall that
 both $N\geq2$ and $k\geq2$). In the first
 case we have $v_{\rho,\sigma}(L)=N\sigma> kN\geq N+k>N-p+k>\rho+\sigma$.
 If $p\geq0$ then
  $v_{\rho,\sigma}(L)=N\sigma=Nk>N+k\geq\rho+\sigma$. The strict inequality
  is a consequence of the fact that $max(N,k)\geq3$.
 \qed

\ble{3.113}Let $L$ act on $\theta$ nilpotently. The polynomial $f$
has the form \beq
 f= (y^r-x)^k, \quad r>1. \label{3.6a}
 \eeq
\ele
          \proof If the term with highest power $k$ in
          \eqref{3.333} is 1 then   $f$ has the form
$f=y^n(y^r-\lambda x)$. In the case when $k\geq2$
           from  \leref{3.112} we know that
$f=y^n(y^r-\lambda x)^k$ or $f=(y-\lambda x)^\a(y-\mu x)^\b $,
$\a+\b=k$, i.e. we have one of the cases  b), c) or d) as $f^s=
g^r$ is impossible. By applying the automorphism
$\Phi_{1,(\mu)^{-1}}^{'}$ of $A_1$ (here $\Phi_{r,-\mu}^{'}(x)=x,
\quad \Phi_{r,\mu}^{'}(y)=y+\mu x^r$)
 to $f$ and to $x^l$ we reduce the last case to the previous
ones, i.e. we assume that $f=y^n(y^r-\lambda x)^k$, where $n\geq1$
and $k\geq1$. Without loss of generality we can assume that
$\lambda=1$.
  Our goal will be to show that if $n\geq1$ and  $k\geq1$ then
$ad_L^s $ cannot be zero for any $s\in \Nset$. We will show that
at least some of the terms with highest weight will be preserved.
Suppose that $n\geq 1, \quad k\geq 1$. First we continue the
automorphisms  $\Phi_{1,r}$ from $A_1$ to its skew-field. Notice
that they preserve the filtration. For this reason we are going to
work only with the homogeneous part as the the rest of the terms
have no impact on it. Apply the automorphism $\Phi_{r,1}$ of $A_1$
to $f$ and to $x^l$. Using
 that $\Phi_{r,1}(\p)=\p$ and $\Phi_{r,1}(x)=x+\p^r$ we obtain

 $$
\Phi_{r,1}(\p^n(\p^r-x)^k)= (-1)^k\p^n x^k , $$

 $$
\Phi_{r,1}(x^l)= (x+\p^r)^l. $$
           Consider now $\ad_{(\p^n x^k)}^s (x+\p^r)^l$. Write

$$ (x+\p^r)^l = \sum c_j^l x^j \p^{r(l-j)} + \ldots, $$
      where by
$\ldots$ we denote the lower weight terms. Then by linearity

 \beq
   ad_{(\p^n x^k)}^s (x+\p^r)^l =
    \sum_{j=0}^l c_j^l ad_{(\p^n x^k)}^s
( x^j \p^{r(l-j)})+
   \ldots  \label{3.8a}
\eeq
 Let $k\geq n$. Simple computation gives that
 $$
 ad_{\p^n x^k}^s (x^l)= [\prod_{j=0}^{s-1}[nl + (k-n)j]\p^{s(n-1)}]
 x^{l+ s(k-1)} + \ldots
 $$
As $n\geq 1,\quad l\geq 1 , \quad k-n\geq 0$ the coefficient at
the term of highest power in $x$ is positive for any $s\geq 1$,
which shows that \eqref{3.8a} cannot be zero for any $s$. Now
suppose that $n\geq k\geq 1$. Consider

$$ ad_{\p^n x^k}^s (\p^{lr})= [(-1)^s \prod_{j=0}^{s-1}[lrk +
(n-k)j]] \p^{s(n-1)+lr}x^{s(k-1)} + \ldots $$
     By the same argument the coefficient at the
 highest power in $\p$ is not zero for any
$s$. This shows that either $n=0$ or $k=0$. But from the assumption
\eqref{3.1} it follows that $k$ cannot be zero.
\qed
   \bre{3.1} Note that the above result gives a normal form for the
leading terms of all bispectral operators with increasing
coefficients of any order.
        \ere

Now assume that the order $N$ of $L$ is a prime number. This gives
that $k=1,\quad N=r$. Using \leref{3.113} we obtain that:
\ble{3.5} The operator L has the form \beq
 L= \p^N + \sum_{j=1}^{N-2}a_j \p^j -x +
\sum_{j=0}^{N-2}W_j(x)\p^j, \label{3.9}
 \eeq
 where
$\lim_{x\rightarrow\infty} W_j(x)=0$ and $a_j \in \Cset$.
\ele
We will call the operator

$$ A=\p^N + \sum_{j=1}^{N-2}a_j \p^j -x $$ the {\em principal
part} of $L$. Following the terminology
 of \cite{BHY1} $A$ is the (generalized) Airy operator.

\subsection{ Airy PDO's}
Let
\beq
 A=\p^N + \sum_{j=1}^{N-2}a_j \p^j - x \label{4.1}
\eeq
        be the generalized Airy operator.
   Our aim here is to develop a calculus of pseudo-differential
   operators written in terms of inverse powers of Airy operators
   in complete analogy with the standard one, described in
   sect.1.1.
         All the results of the
section are obtained for any order $N$, i.e. without assuming that
the order is prime. Let $\Phi(x)$ be a nonzero
 function in $Ker A$, i.e.
 \beq
 A\Phi(x)=0. \label{4.2}
 \eeq
 Then the function $\Psi(x,z)=\phi(x+z)$ satisfies the equations:

 \beqa
  &&A(x,\p_x)\Psi(x,z)=z\Psi(x,z),\label{4.3}\\
 &&A(z,\p_z)\Psi(x,z)=x\Psi(x,z).  \label{4.4}
 \eeqa
 Obviously the function $\Psi(x,z)$ satisfies also the equation

 \beq
 \p_x \Psi(x,z)= \p_z \Psi(x,z). \label{4.5}
\eeq
 The equations \eqref{4.3}-\eqref{4.5} define an anti-involution
 $b$ on the Weyl algebra $A_1$ (see \cite{BHY2, W1}),
 acting on the generators $A,\p_x$ of $A_1$ by

 \beqa
 b(A(x,\p_x))=z,\\
 b(\p_x)=\p_z. \label{4.6}
 \eeqa
 From \eqref{4.6} one easily finds that
 \beq
 b(x)=A(z,\p_z), \label{4.7}
 \eeq
 which would be used later.

 Now define the algebra $\mathcal{B}_1$ of pseudo-differential
 operators of the type:
 \beq
 P(x,\p_x)= \sum_{j=-m}^{\infty}a_j(x,\p_x)A^{-j}, \label{4.8}
\eeq
with operator coefficients of the form:
\beq
 a_j=\sum_{k=0}^{N-1}\a_{j,k}(x)\p^k, \label{4.9}
 \eeq
where the functions $\a_{j,k}(x)$ are formal Laurent series:

\beq
 \a_{j,k}(x)= \sum_{s=r}^{\infty}\b_{j,k}^{(s)}x^{-s}
\label{4.10}
 \eeq
  and the index $r$ depends only on $P$ ({\em but not on $j$!}).
 Then
  the anti-automorphism $b$ can be continued on $\mathcal{B}_1$ as

\beq b(P)= \sum_{j=-m}^{\infty}z^{-j}\sum_{k=0}^{N-1}
\sum_{s=r}^{\infty}\b_{j,k}^{(s)}\p_z^k A^{-s}   \\
     =\sum_{s=r}^{\infty}b_s(z,\p_z) A^{-s}(z). \label{4.11}
\eeq

Introduce the "wave operator"  $K$ as follows:

\beq
 K=1+\sum_{j=1}^{\infty}m_j(x,\p_x)A^{-j}, \label{4.12}
  \eeq
where
 \beq
m_j(x,\p_x)=\sum_{k=0}^{N-1}\a_{j,k}(x)\p_x^k. \label{4.12a}
 \eeq
   Let $L$ be a differential operator of the form:

 \beq
 L=A^l+V_{l-1}A^{l-1}+\ldots, \label{4.13}
 \eeq
 where
 \beq
 V_j(x,\p)=\sum_{k=0}^{N-1}V_{j,k}(x)\p^k. \label{4.14}
 \eeq
 Then one can find a wave operator (not unique) $K$ of the form
\eqref{4.12} so that

   \beq
    L=KA^lK^{-1}. \label{4.15}
    \eeq
    The coefficients
  $\a_{j,k}(x)$  from \eqref{4.13} of the expansion
 of $K$ can be found by induction from the equation:
 \beq
 LK=KA^l \label{4.16}
 \eeq
 Multiplying the above equation from the right
  by $A,A^2, \ldots$ one computes  the coefficients
   $\a_{j,k}(x)$ in the expansion
\eqref{4.12} - \eqref{4.13} of
$K$ in terms of the functions $V_{j,k}$. We would
 particularly
be  interested in the case $l=1$.

 In what follows we assume that the operator $L$ is
 bispectral, it
 satisfies equation of the form \eqref{1.1}, together
 with an
 equation of the form \eqref{1.2}. In that case as in
 \cite{DG,W1} one can prove the following lemma.

 \ble{4.1}
  The coefficients $a_{j,k}$ of the operator $K$ are
 rational functions.
 \ele
 \proof   We mimic the well known proof (see \cite{DG,W1}).
 Write

 \beq
(\ad_L)^m(\theta)=0 .      \label{4.17}
  \eeq
  This is equivalent to
  \beq
  (ad(A^l))^m(K^{-1} \theta K). \label{4.18}
  \eeq
  Put
  \beq
  \Theta=K^{-1}\theta K = \sum_{j=0}^\infty\theta_j A^{-j},
  \label{4.19}
  \eeq
  where

  \ben
  \theta_j = \sum_{k=0}^{N-1} \theta_{j,k} \p^k.
  \een
  This gives
  \ben
  (adA^l)^m (\theta_j)=0 .
  \een
The leading terms of the above equation give

\ben
\theta_{j,N-1}^{(m)}\p^{m(Nl-1)+N-1}+
( \theta_{j,N-2}^{(m)}\ldots)\p^{m(Nl-1)+N-2}+\ldots +\\
( \theta_{j,0}^{(m)}\ldots)\p^{m(Nl-1)}+
 \ldots =0
 \een
 Here in the brackets containing the
 coefficients at $\p^{m(Nl-1)+N-s},\quad s=2,\ldots,N$ the dots
 after $\theta_{j,N-s}$ denote expressions of
 derivatives of $\theta_{j,N-r}, r<s$
 of order not lower than $m$. Then obviously by induction
 we get that all $\theta_{j,n-1}^{(m)}\equiv0, \quad n=1,\ldots, N$,
  which shows that
 they are polynomials of degree $d \leq m$. Then the computation of the
 coefficients $a_j(x)$ of $K$ is performed as in \cite{DG,W1}
(see also sect.2). We see that they are  rational functions.
   \qed

     \subsection{Proof of the main
     theorem for operators with increasing coefficients}

Let $L$ be an operator of order $N$ with Airy principal part
    \ben
    A=\p^N + a_{N-2}\p^{N-2}+\ldots+a_1\p - x,
    \een
    i.e.
     \ben
  L=A + \sum_{j=0}^{N-2}V_j(x)\p^j =A+ V(x,\p_x)
  \een
  and
      \ben
       \lim_{x\rightarrow\infty}V_j(x)=0.
    \een

 Using the techniques of the previous subsection we
present $L$ in the form
\begin{equation}\label{5.3}
L=A+V = KAK^{-1}
\end{equation}

Our goal will be to show that bispectrality, and in particular -
the rationality of the coefficients $\a_{j,k}(x)$ in the expansion
\eqref{4.12} and \eqref{4.12a} of $K$, implies that the
perturbation $V(x,\p_x)\equiv0$, which is equivalent to
$\a_{j,k}\equiv0$ for all $j,k$. But first we need some notation
and auxiliary results.

             From the equation

\beq
L K= K A, \label{5.5}
\eeq
      we
 can compute recursively the coefficients $m_j$ of
 the operator $K$.
 For this we will need some formulas
to compare the coefficients of the two sides of \eqref{5.5}.
Introduce the operators $b_j$,  $c_j$, $U_j$, $W_j$, $j= 1,2,
\ldots$ by:

      \beqa
      && [A,m_j]= b_jA+c_j; \label{5.6}\hfill\\
       &&  V(x,\p)m_j= U_jA+W_j.\label{5.7}\hfill
       \eeqa


We will need to order the monomials in $m$ and related to it
expressions as follows:

\bde{5.1} We say that the monomial $x^{r_1}\p^{k_1}$ is of higher
order than the monomial  $x^{r_2}\p^{k_2}$  if $r_1> r_2$ or $r_1=
r_2$ and $k_1> k_2$. We will call the number $r$ the {height} of
$m$, if the highest order term of $m$ is of the type $c.x^{r}\p^k,
\quad c\neq0$. We denote this number by $ht(m)$.
       \ede
      In other words we use lexicographic ordering in the set of
the monomials $x^r\p^k$ but the order is only the power of $x$.

 In what follows we are going to use the abbreviation $l.o.t.$
 (for {\em lower
order terms}) compared to some operator $m$ with the  meaning that
they are lower than at least one of the terms in $m$. We will need
 also the following lemma:

\ble{5.3} In the above formulas we have

(i) \beq b_j=\sum_{k=1}^{N-1}N\a_{j,k}^{'}\p^{k-1}+
       l.o.t. \label{5.8}
\eeq

\beq c_j = N\a_{j,0}^{'}\p^{N-1}+
 \sum_{k=1}^{N-1}(xN\a_{j,k}^{'}+ k\a_{j,k})\p^{k-1}
 + l.o.t.\label{5.9}
\eeq

(ii)  $ht( c_j)= ht(b_j)+1$.
    \ele
\proof The proof is straightforward computation. To avoid two
indices we will suppress the dependence on $j$ (it is irrelevant
at that moment). We have

    \beqa
  && A\circ m = m(\sum_{k=1}^{N-1}a_k\p^k) -x m +
    \sum_{k=0}^{N-1}N\a_{k}^{'}\p^{N+k-1} + l.o.t. \label{5.2p}\hfill\\
   && m\circ A =  m(\sum_{k=1}^{N-1}a_k\p^k) -x m +
   -\sum_{k=0}^{N-1}k\a_{k}\p^{k-1}\label{5.3p}. \hfill
      \eeqa
  Subtracting \eqref{5.3p} from  \eqref{5.2p} we get

  \beq
   [A,m] = \sum_{k=0}^{N-1}N\a_{k}^{'}\p^{N+k-1} +
    \sum_{k=0}^{N-1} \a_{k}\p^{k-1}+ l.o.t. \label{5.4p}.
  \eeq
Split the first sum into two parts as follows. One of them
contains derivatives from $N$ to $2N-2$; the second will contain
the rest of the them. Then we have for the first part

  \ben
     \sum_{k=0}^{N-1}N\a_{k}^{'}\p^{N+k-1}=
     (\sum_{k=1}^{N-1}N\a_{k}^{'}\p^{k-1})\p^N
     \een
Next use the identity $\p^N = A- \sum_{k=1}^{N-2}a_k +x$ to get
  \ben
     N\a_{0}\p^{N-1} + \sum_{k=1}^{N-1}N\a_{k}^{'}\p^{N+k-1}=\\
     N\a_{0}\p^{N-1} + ( \sum_{k=1}^{N-1}N\a_{k}^{'}\p^{k-1})A +
     x( \sum_{k=1}^{N-1}N\a_{k}^{'}\p^{k-1})+ \ldots
       \een
where the dots represent terms with derivatives of $\a_{k}$. Then
we repeat the same procedure to all terms (including the ones in
l.o.t. from \eqref{5.4p}) containing $\p^k$ with  $k \geq N$.
After finite number of steps we get \eqref{5.8} and \eqref{5.9}.
The second part of the lemma follows immediately from the first
one. \qed

       \ble{5.2}
    (i) The equation \eqref{5.5}  is equivalent to the equations:

  \beqa
     && b_1+ V + U_1=0 \label{5.10a} \hfill\\
     && b_{j+1}+c_j+W_{j}+U_{j+1}=0,
    \quad j=1,\ldots \label{5.10}\hfill
       \eeqa
  (ii) The coefficients of $V$ and $b_1$ behave at infinity as $x^{-2}$
        \ele

        \proof  The first part is simply comparing the coefficients.
        Indeed, writing in detail \eqref{5.5} we get

        $$ A+Am_1A^{-1} +\ldots + V + Vm_1A^{-1} +\ldots=
        A + m_1 + \ldots $$
        Using \eqref{5.6} and \eqref{5.7} we can simplify the last
        equation to
        $$V+b_1 +U_1 + \ldots=0,$$
       where $\ldots$ denote the purely pseudo-differential part.
       This gives \eqref{5.10a}. Multiplying \eqref{5.5} by $A$,
       $A^2$, etc. from the right and arguing in the same manner we get
       \eqref{5.10}.

       To prove the second part we use \eqref{5.8} with $j=1$ and
       \eqref{5.10a}. Notice that the leading terms of $b_1$ are
       derivatives of a rational functions. Being equal to the
       leading terms of $V$ they vanish. Hence they vanish at
       least of order $x^{-2}$. \qed
    \ble{5.4}
    The following inequalities hold:
    \beqa
     &&ht(U_j) \leq ht(b_j) - 1,  \label{5.7p}  \hfill\\
          &&ht(W_j \leq ht(c_j) -1.  \label{5.8p} \hfill
    \eeqa
    \ele
\proof The proof is similar to that of \leref{5.3}. Using
\eqref{5.7} we obtain

   \ben
   Vm_j = \sum_{k=1}^{2N-1}( \sum_{s=k}^{N-1}V_s\a_{j,k-s})\p^k
        = \sum_{k=0}^{2N-1}\widetilde V_{j,k}\p^k
  \een
  For $k=N,\ldots,N-1$ put $\tilde{V_{j,k}}= \tilde{U_{j,k-N}}$.
  As above split the sum into two parts, the first one containing
  the terms with $\p^k$, $k \geq N$:

  \ben
 Vm_j=(\sum_{k=0}^{N-1}\widetilde U_{j,k}\p^k)\p^N +
   \sum_{k=0}^{N-1}\widetilde V_{j,k}\p^k
      \een
  Again use the identity $\p^N=A - \sum_{k=1}^{N-2}a_k +x$
  several times  to get the first sum in the form:

       \ben
     ( \sum_{k=0}^{N-1}\widetilde U_{j,k}\p^k + l.o.t.)A +
      x \sum_{k=0}^{N-1}\widetilde U_{j,k}\p^k +l.o.t.
      \een
Then obviously we have:

      \beqa
  && U_j=  \sum_{k=0}^{N-1}\widetilde U_{j,k}\p^k + l.o.t.
   \label{5.9p}\hfill\\
  && W_j =  \sum_{k=0}^{N-1}\widetilde V_{j,k}\p^k +
   x \sum_{k=0}^{N-1}\widetilde U_{j,k}\p^k +l.o.t.
   \label{5.10p}\hfill
     \eeqa
  Now using that the order at infinity of $V$ is $x^{-2}$ we get
  that $ht(U_j)\leq ht(m)-2$, $ht(W_j)\leq ht(m_j)-1$. From
  the last inequalities we get \eqref{5.7p} and \eqref{5.8p}. \qed

\vskip5pt

Now we are ready to finish the proof of the \thref{1.3}.

\proof of \thref{1.3}
 We recall that we have to show that $V\equiv0$.  Assume
that some of the coefficients $V_j$ are not zero. Then we shall
compute the leading terms of the operators $b_j$ recursively using
\eqref{5.10} and \leref{5.2} and taking into account the estimates
\eqref{5.7p} and \eqref{5.8p}
 First notice that that the highest
 order term in $b_1$ is of the type
 $\a_1 x^{s_1} \p^k, \a_1\neq 0$,
 with $k<N$ and $s_1<0$. Suppose that
  the highest order term in $b_{j}$ is
  $\a_j x^{s_j} \p^k, \a_j\neq 0$. Then the highest order
term in $b_{j+1}$ is computed, using \eqref{5.10}, \eqref{5.8} and
\eqref{5.9} to be $\a_{j+1}x^{s_j+1}\p^k$ with  $\a_{j+1}= -
\a_j(N(s_j+1) + k)/N(s_j+1)$. Having in mind $k<N$ we get that
$\a_{j+1}\neq 0$. After a finite number of steps we will get that
for some $j$ the corresponding $s_j=-1$. But this contradicts the
fact that the highest order term of $b_j$ is a derivative of a
rational function. \qed


\sectionnew{Final remarks on the proof and comments}

\subsection{Proof of Theorem 0.1}
Essentially we already have performed the proof of the main
theorem. We just have to notice that when the order of $L$ is
prime and there are coefficients increasing at infinity
\leref{3.5} and \thref{1.3} give that the operator is Airy and
hence bispectral.

If the coefficients of $L$ are bounded (at infinity) and the order
is prime, then using the fact the the rank divides the order we
get that either the rank of $L$ is $1$ or it is equal to its
order. The latter case is treated in \thref{1.2}. If the rank is
one then this is the main result of \cite{W1}. Finally the inverse
part, i.e. that all the operators listed in \thref{1.1} are
bispectral is the main result of \cite{BHY1}.\qed

\subsection{Comments}

Here I would like to make some speculations on eventual
continuation of the classification. It seems to me that the
methods of \cite{BHY3} (see also \cite{H1} for more details) will
be enough to construct all bispectral operators.  Assuming that
then the classification should be: 1) find all "basic" bispectral
operators; 2)show that Darboux transforms reduce any bispectral
operator to a "basic" one.

Having in mind the constructions in \cite{DG, BHY1, KRo, W1} it
seems natural to consider basic those operators  that have as few
singularities as possible and generate their centralizers. Then in
view of the main result of \cite{H2} one class of operators that
certainly should be considered as "basic" is the class of
bispectral operators $L$ in the Weyl algebra that together with
some other operator $Q$ satisfy the "canonical commutation
relation" (CCR):

\beq
 [L,Q]=1 \label{13.1}
   \eeq
 More precisely they are basic because according to \cite{H2}
  all bispectral operators
 in the Weyl algebra are simply polynomials in operators $L$ that satisfy
 \eqref{13.1}.
   It is tempting to believe that the CCR is enouhg for bispectrality:

   \bcon{13.1}
If the operators $L$ and $Q$ satisfy the CCR \eqref{13.1} then
they are bispectral.
   \econ
   This conjecture seems to be a difficult one
   as it is easily shown (cf.\cite{H2}) to be equivalent
   to the famous conjecture of Dixmier-Kirillov:

   \bcon{7.6}
  If the operators $L,Q$ satisfy the CCR \eqref{13.1}
   then they generate the Weyl algebra $A_1$. In
  other words, any endomorphism of the Weyl algebra is an automorphism.
          \econ

The Bessel operators as well as other examples from \cite{BHY3}
show that one needs also to add to the list of the basic operators
$L$ the ones that together with some other operator $Q$ satisfy
the following "string" equation

 \beq
[L,Q]=L \label{13.2}
          \eeq
Our, maybe insufficient experience, suggests that the above two
classes contain all basic bispectral operators. A more precise
conjecture is the following one:

\bcon{13.2}
   Every bispectral operator is a polynomial Darboux transformation of a
   bispectral operator satisfying either \eqref{13.1} or
   \eqref{13.2}.
     \econ

     To make the classification of the basic bispectral operators
     more explicit  introduce the following notation. Denote by
     $B_{\a}$ the algebra spanned by a Bessel operator $L_{\a}$,
     $x^N$ and $D$. Then

\bcon{13.3}
    (1) A bispectral operator satisfying \eqref{13.1}  belongs the Weyl
    algebra.  (2) A bispectral operator satisfying \eqref{13.2}
    belongs to one of the algebras $B_{\a}$.
  \econ

Some progress could be achieved if one finds   analogs for
$B_{\a}$ of the results from \cite{H2}. One difficulty would be to
extend Dixmier results for these algebras.

 All the above
conjectures aim to a complete classification of bispectral
operators. A more modest goal is to find the bispectral operators
with some properties. Here is a conjecture in this direction.

      \bcon{7.1} An operator L with bounded at infinity coefficients is
bispectral if and only if it is a polynomial Darboux
transformation of a  Bessel operator.
       \econ
        This conjecture seems natural in view
of the "if" part, obtained in \cite{BHY1}. It will be very useful
either to give a proof of Wilson's result about rank one operators
without using algebraic-geometric arguments or modify his proof in
higher rank situation. Despite of the many interesting results for
higher rank solutions of KP-hierarchy I have not found a
construction suitable for our purposes. In any case the above
conjecture seems to me to be within the reach of the existing
tools unlike the previous ones.



\renewcommand{\em}{\textrm}
\begin{small}
\renewcommand{\refname}{ {\flushleft\normalsize\bf{References}} }
    
\end{small}
\end{document}